

\input phyzzx
\rightline{UTS-DFT-93-12}
\vfill
\title{Kac-Moody structure of chiral gravity in the light-cone gauge}
\vfill
\author{A.Smailagic\foot{E-mail address: ANAIS@ITSICTP.BITNET}}
\bigskip
\address{International Center for Theoretical Physics, Trieste\break
Strada Costiera 11, 34014, Trieste, Italy}
\andauthor{Euro Spallucci\foot{E-mail address:SPALLUCCI@TRIESTE.INFN.IT}}
\bigskip
\address{Dipartimento di Fisica Teorica\break
Universit\`a di Trieste,\break
INFN, Sezione di Trieste,\break
Trieste, Italy 34014}
\vfill
\submit{Phys.Lett.B}
\vfill\endpage

\abstract
We study the residual symmetry $SL(2,R)\otimes U(1)$
of the chiral gravity in the light-cone gauge. Quantum gravitational
effects renormalize the Kac-Moody central charge and introduce, through the
Lorentz anomaly, an arbitrary parameter. Due to the presence of this free
parameter the Kac-Moody central charge has no forbidden range of values,
and the strong gravity regime is open to investigations.

\vfill\endpage

\hfuzz 30pt
\REF\rs{ J.Ambiorn, B.Durhuus, J.Fr\"olich
\journal Nucl.Phys. &B257[FS14] (85) 433;\hfill\break
F.David
\journal Nucl.Phys. &B257[FS14] (85) 45;\hfill\break
V.Kazakov, I.K. Kostov, A.A.Migdal
\journal Phys.Lett. &B157 (85) 295;\hfill\break
V.Kazakov, I.K. Kostov, A.A.Migdal
\journal Nucl.Phys. &B311 (90) 127.}
\REF\mat{ D.J.Gross, A.A.Migdal
\journal Phys.Rev.Lett. &64 (90) 717;\hfill\break
E.Brezin, V.Kazakov
\journal Phys.Lett. &236B (90) 144.}
\REF\conf{V.G.Knizhnik, A.M.Polyakov, A.B.Zamolodchikov
\journal Mod.Phys.Lett. &A3 (88) 819.}
\REF\polya{A.M.Polyakov
\journal Mod.Phys.Lett. &A2 (87) 893.}
\REF\pert{S.Ichinose
\journal ``Perturbative approach to 2 dimensional quantum gravity''
&YITP/K-876 (90) {} preprint;\hfill\break
G.W.Delius, M.T.Grisaru, P.Van Nieuwenhuizen
\journal Nucl.Phys. &B389 (93) 25.}
\REF\tse{A.A.Tseytlin
\journal Int. J.Mod.Phys.Lett. &A5 (90) 1833.}
\REF\pol{A.M.Polyakov, A.B.Zamolodchikov
\journal Mod.Phys.Lett. &A3 (88) 1213.}
\REF\vafa{H.Ooguri, C.Vafa
\journal Mod.Phys.Lett. &A5 (90) 1389.}
\REF\anais{A.Smailagic
\journal Phys.Lett. &B195 (87) 213.}
\REF\jr{R.Jackiw, R.Rajaraman
\journal Phys.Rev.Lett. &54 (85) 1219.}
\REF\quant{A.M.Polyakov
\journal in Field, Strings and Critical Phenomena &{} (89)
E.Brezin, J.Zinn-Justin, Les Houches.}
\REF\noi{A.Smailagic, E.Spallucci
\journal Phys.Lett. &B284 (92) 17.}
\REF\reuter{A.H.Chamseddine, M.Reuter,
\journal Nucl.Phys. &B317 (89) 757.}
\REF\lee{K.Li
\journal Phys.Rev. &D34 (86) 2292;\hfill\break
T.Fukuyama, K.Kamimura
\journal Phys.Lett. &B200 (88) 75;\hfill\break
T.Lee
\journal ``~two dimensional chiral quantum gravity~'' &SNUTP-91-54, (92),
{} preprint.}
\REF\berg{T.Berger
\journal ``Fermions in Two (1+1) Dimensional Anomalous Gauge Theories''
&90-084, (90) Desy preprint.}
\REF\leut{H.Leutwyler
\journal Phys.Lett. &B153 (85) 65.}
\REF\meyer{R.C.Myers, V.Periwal
\journal Nucl.Phys. &B397 (93) 239.}
\REF\dk{J.Distler, H.Kaway
\journal Nucl.Phys. &B321 (89) 509;\hfill\break
F.David
\journal Mod.Phys.Lett. &A3 (88) 1651.}
\REF\kl{J.Kim, T.Lee
\journal Phys.Rev. &D42 (90) 2664.}
\REF\oz{Y.Oz, J.Pawelczyk, S.Yankielowicz
\journal Nucl.Phys. &B363 (91) 555.}



\doublespace

\chapter{}

Recently much work has been devoted to the study of quantum two dimensional
gravity both in connection with strings out of critical dimensions and
as a useful ``~laboratory~'' to get insight into realistic four dimensional
gravity.  The advantage of two dimensional gravity is that it is exactly
solvable either in the context of random surfaces theory [\rs], matrix
models [\mat], conformal field
theory [\conf], or in a perturbative approach [\polya,\pert].
The choice of the light-cone gauge is one of the main ingredients in
two dimensional gravity calculations. This gauge choice has at least two
distinctive advantaged: first, light-cone coordinates led
to the discovery of underlying Kac-Moody algebra of residual gauge
symmetry; secondly, it provided
a satisfactory regulator for perturbative calculations [\tse].

The above mentioned approaches displayed a remarkable result: the
renormalized Kac-Moody central charge can attain complex value where
physical information is lost. As a result, the weak gravity
$c\le 1$ and strong gravity $c\ge 25$ regimes (~$c$ denotes the total
number of matter fields~) are separated by a
``~phase transition~'' , which forbids a satisfactory investigation
of the strong gravity domain. A more promising framework is provided
by $N=2$ super-gravity models where there
is no gap at all between the weak and strong gravity regions [\pol].
Unfortunately, $N=2$ models have
unphysical signature $(++--)$, and their physical relevance is
presently unclear,
although some effort has been devoted to provide a physical
justification for them [\vafa].

In this letter, we are going to study a {\it chiral, induced, quantum gravity}
in the light-cone gauge. Our main interest in this model concerns the
presence of anomalies which cannot be removed by
local counter-terms, and therefore naturally incorporate regularization
dependent free parameter [\anais].
Though anomalous, chiral theories can achieve
consistency at least in the case of the chiral
Schwinger model [\jr]. In a similar spirit we shall investigate
the Kac-Moody
structure of chiral gravity in light-cone gauge with the hope that the
presence of additional dynamical (~Lorentz~) degree of freedom may improve,
or possibly avoid forbidden regions of values for renormalized
Kac-Moody central charge.
This hope is sustained by the known fact that the Kac-Moody central charge
corresponding to an abelian symmetry
(~which, in our case, is described by Lorentz local invariance~)
does not get renormalized by quantum effects [\quant]. In this case
all one needs to
calculate is the Kac-Moody central charge corresponding to residual
diffeomorphisms invariance
(~as in the case of non-chiral gravity~) by taking into account the
additional Lorentz contribution.

The dynamics of
chiral induced gravity can be encoded into the symmetric action [\noi]
$$
S=
{1\over 2}\int d^2x\sqrt{-g}\left[\widetilde R{1\over\nabla^2}\widetilde R+
a'\omega^2\right]\ ,\quad
\widetilde R=\alpha R+\beta \nabla\omega\ ,
\eqn\uno
$$
where: $\alpha$ and $\beta$ are constants related to the sum and the difference
$n_{\pm}$
of the number of left and right chirality components of matter fermions,
in a manner which we shall describe later; $\nabla_\mu$
is the Christoffel generally covariant derivative.

The action \uno\ displays the dependence from the spin connection
$\omega_\mu$, acting as the gauge field of the local Lorentz  symmetry.
The non-local term in $\omega_\mu$ is the origin of the
Lorentz anomaly, which cannot be removed by any choice of the coefficient
$a'$ in front of the local term $\omega^2$. Strictly speaking, eq.\uno\
describes a whole family of actions, each member of the family being labelled
a different value of $a'$ corresponding to a different choice of the
regularization scheme. Finally, the metric tensor $g_{\mu\nu}$ is actually
a composite object, built up form the more fundamental zweibein field
$e^a{}_\mu(x)$. Accordingly, the
energy-momentum current should be properly defined as the response of
the action under zweibein variation. However, we find more comfortable
to work with the energy-momentum tensor defined as
$$\eqalign{
T_{\mu\nu}&\equiv -{e^a{}_\mu\over\sqrt{-g}}{\delta S\over\delta e^a{}_\nu }\cr
&=2\alpha\nabla_\mu\nabla_\nu\phi-\nabla_\mu\phi\nabla_\nu\phi
+\beta\left(\omega_\mu\nabla_\nu\phi+\omega_\nu\nabla_\mu\phi\right)\cr
&-a'\omega_\mu\omega_\nu-g_{\mu\nu}\left(2\alpha\nabla^2\phi-{1\over 2}
\nabla^\rho\phi\nabla_\rho\phi+\beta\omega^\rho\nabla_\rho\phi-{a'\over 2}
\omega^2\right)\cr
&-2\beta\left[\epsilon_{\mu\nu}\nabla^2\phi+\epsilon_{\nu\rho}\nabla^\rho
\nabla_\mu\phi\right]+2a' \left[\epsilon_{\mu\nu}\nabla\omega+
\epsilon_{\nu\rho}\nabla^\rho\omega_\mu\right]\ ,
\cr}
\eqn\tre
$$
where we introduced the auxiliary, scalar, field $\phi$, which allows to
write the action \uno\ in a local form, and is related to $\widetilde R$ by
$\nabla^2 \phi=\widetilde R$.

We shall be working in the light-cone gauge where the metric reads
$$
ds^2=dx^+dx^- +h_{++}dx^+dx^+\ ,
\eqn\cinque
$$
and the spin-connection components can be written in terms of $h_{++}$,
and the Lorentz degree of freedom $L$, as
$$
\eqalign{
&\omega_+=\left[\partial_+L+2\partial_-h_{++}\right]\ ,\cr
&\omega_-=\partial_-L\ .\cr}
\eqn\sei
$$
Furthermore, $L$ is defined in terms of the zweibein components through
$$
e^L=e^{\widehat{+}}{}_+e^-{}_{\widehat{-}}\ .
\eqn\sette
$$
Lorentz indices in eq.\sette\  are denoted by a hat to distinguish them
from (~un-hatted~) world indices.

Since the action \uno\ is only diffeomorfism invariant,
general covariance is the only symmetry one can use to eliminate redundant
degrees of freedom. We choose to eliminate $h_{--}$ and $S={1\over 2}
(h_{+-}+h_{-+})$, while $h_{++}$ and $L$(Lorentz degree of freedom) remain
as dynamical variables. Therefore, various components of $T_{\mu\nu}$ have
distinct role. $T_{--}$ and $\epsilon^{\mu\nu}T_{\mu\nu}$ couple to
dynamical degrees of freedom and will generate equations of motion governing
dynamics of chiral gravity. These components are given by
$$
T_{--}(\phi)=\left[2(\alpha+\beta)\partial_-^2\phi
-\left(\partial_-\phi\right)^2
+2\beta \omega_-\partial_-\phi-2a'\partial_-\omega_-
-a'\omega_-^2\right]\ ,
\eqn\otto
$$
and
$$
\epsilon^{\mu\nu}T_{\mu\nu}=-2\alpha\beta R+2\left(a'-\beta^2\right)
\nabla\omega\ .
\eqn\nove
$$
Resulting equations of motion are
$$
-\left[\left(\alpha\pm\beta\right)^2
+\left(a'-\beta^2\right)\left(1\mp {\alpha\beta\over
a'-\beta^2}\right)^2\right]
\partial_-^3 h_{++}-\alpha\beta\partial_-^2 A_+=0\ ,
\eqn\novenove
$$
$$\eqalign{
(a'-\beta^2)\partial_- A_+&=0\ ,\cr
(a'-\beta^2)&\ne 0\ ,\cr}
\eqn\dieci
$$
where we have introduced convenient redefinition
$$
A_+=D_+L +\left(1-{\alpha\beta\over a'-\beta^2}\right)\partial_-h_{++}
\eqn\undici
$$
which gives \dieci\ a simple looking form,
and decouples fields in the lagrangian
\uno\ which then written in the light-cone gauge is:
$$
L=\left[\left(\alpha\pm\beta\right)^2+
\left(a'-\beta^2\right)\left(1\mp {\alpha\beta\over a'-\beta^2}\right)^2\right]
\partial_-^2 h_{++}{1\over D_+}\partial_-h_{++}-
\left(a'-\beta^2\right)\partial_-A_+{1\over D_+}A_+\ .
\eqn\unouno
$$

The new derivative $D_+$
is defined as $\displaystyle{D_+ L=\partial_+ L-h_{++}\partial_-L}$. On the
other hand, components of the energy-momentum tensor $T_{++}$ and $T_{+-}$
that couple to the gauge degrees of freedom $h_{--}$ and $S$(=Weyl degree
of freedom) are the generators of the residual symmetry of
the invariant line element \tre\ in the light-cone gauge [\reuter],
and weakly vanishing condition must be imposed in order
to preserve the residual symmetries at the quantum level.
It is worth mentioning that the Lorentz symmetry (~or its absence~) has
no influence on the line element and serves only to induce dynamics
for the Lorentz degree of freedom through \nove.
These components of the energy-momentum tensor are
$$
T_{++}=
\left[
-\left(\partial_-h_{++}\right)^2+2h_{++}\partial_-^2h_{++}
-2Q_1\partial_-\partial_+h_{++}\right]
+\left[A_+^2 -2Q_2\partial_+A_+\right]
\eqn\dodici
$$

where we have conveniently rescaled fields according to
$$\eqalign{
&h_{++}\rightarrow \left[-\left(\alpha\pm\beta\right)^2-
\left(a'-\beta^2\right)\left[1\mp {\alpha\beta\over a'-\beta^2}\right]^2
\right]^{1/2}h_{++}\ ,\cr
&A_+\rightarrow \left(a'-\beta^2\right)^{1/2}A_+\ ,\cr}
\eqn\tredici
$$
with the constants $Q_1, Q_2$ defined as
$$\eqalign{
Q_1&\equiv \left[-\left(\alpha\pm\beta\right)^2-
\left(a'-\beta^2\right)\left[1\mp {\alpha\beta\over a'-\beta^2}\right]^2
\right]^{1/2}\ ,\cr
Q_2&\equiv \left(a'-\beta^2\right)^{1/2}\left(1+{
\alpha\beta\over a'-\beta^2}\right)\ .\cr}
\eqn\quattordici
$$
We would reasonably expect the rescalings \tredici\ to be real in order to
preserve the physical character of the corresponding fields. But, this
condition is not guaranteed by eqs.\tredici\ and must be imposed as a
constrain on the parameters. Reality of the first rescaling in eqs.\tredici\
gives
$$
-{1\over 4}\left(|\alpha+\beta|+|\alpha-\beta|\right)^2< a'-\beta^2<
-{1\over 4}\left(|\alpha+\beta|-|\alpha-\beta|\right)^2\ ,
\eqn\zero
$$
and, unavoidably, leads to a purely imaginary rescaling of $A_+$.
As a consequence, the kinetic term of $A_+$ flips from the
correct to the ``~wrong~'' sign, and the rescaled field becomes a
ghost-like object. Alternatively, one could maintain $A_+$ real at the
expense of assigning $h_{++}$ a ghost-like character.
In what follows we shall adhere to the former choice. The impossibility
of having simultaneously both $h_{++}$ and $A_+$  in the physical sector, is
the light-cone gauge analogue of a similar result obtained in the conformal
gauge, where, in order to have a physical Liouville field, the Lorentz degree
of
freedom must belong to the ghost-like sector [\lee].

As we mentioned earlier, constants $\alpha,\beta$ are related to the
number of left and right chirality component of matter in the following
way [\berg,\noi]
$$\eqalign{
&\alpha^2+\beta^2={N\over 96\pi}\cr
&\alpha\beta={\Delta N\over 192\pi}\cr
&\hat a=192\pi\left(a'-\beta^2\right)\cr
&N=n_+ + n_-\cr
&\Delta N=n_+ - n_-\cr
&\left(\alpha\pm\beta\right)^2={n_\pm\over 48\pi}\cr}
\eqn\quindici
$$
Eqs.\quindici\ give the link among the parameter in our symmetric action
\uno\ and the parameters, in the asymmetric
action for chiral gravity, that usually appears in the literature. Therefore,
eqs.\quindici\ provide the translation code among our formulae
and those in other papers [\berg,\leut].

To complete our calculation, it is further necessary to account for the
ghosts corresponding to the
light-cone gauge choice, as well as for the non-trivial Jacobian following
from the field redefinition \undici. It is possible to give a unique
formula containing all these contributions by noticing that either ghosts or
the Jacobian are expressed in terms of the generalized derivative
$$
D_{(q)\,+}=\partial_+ -h_{++}\partial_- -q\partial_- h_{++}
\eqn\sedici
$$
where $q$ is the Lorentz weight of the field, the operator $D_{(q)\,+}$
is acting on. Therefore, one can find the following result
$$
\left({\rm det}D_{(q)\,+}\right)^{\left(-n\right)^{s+1}}=\exp\left[i
{(-1)^s n\over 24\pi}
(6q^2-6q+1)\int d^2x\, \partial_-^2 h_{++}{1\over D_+}\partial_-h_{++}\right]
\eqn\diciasette
$$
where $s$ denotes the statistics of the involved fields, and $n$ is either $1$
or $1/2$ depending whether field variables are complex or real.
$T_{++}$ in eq.\dodici\  does not contain quantum gravitational contributions.
Its Virasoro central charge is given by
$$\eqalign{
&c_{\rm grav.}=28-n_-\ ,\cr
&c_{\rm grav.}=2+48\pi\left(Q_1^2+Q_2^2\right)\ .\cr}
\eqn\diciannove
$$
 This the same result obtained in the
conformal gauge [\conf] with the DDK argument of the vanishing of the
total central charge. It is worth mentioning that the {\it
Virasoro central charge is independent of the free parameter} due to
the explicit
cancellation between $Q_1^2$ and $Q_2^2$.
The solutions of the equations of motion
\nove,\dieci\ are:
$$\eqalign{
h_{++}(x)&=J_{++}(x^+)-2x^- J_+(x^+)+\left(x^-\right)^2 J_0(x^+)\ ,\cr
A_+(x)&=\widetilde J_+(x^+)\ .\cr}
\eqn\venti
$$
With the help of the anomaly equations \otto,\nove\ it is possible to
obtain Ward identities relating various multipoint functions [\polya] which
in our case read
$$\eqalign{
\biggl\langle h_{++}(x)h_{++}(x_1)\dots &h_{++}(x_n)\biggr\rangle=
\sum_{i=1}^n\Bigl\{
4\pi Q^2_1\left({x^- -x_i^-\over x^+ -x_i^+}\right)^2
\biggl\langle h_{++}(x)\dots \widehat{h_{++}}(x_i)\dots h_{++}(x_n)
\biggr\rangle +\cr
&\left[{(x^- -x_i^-)^2\over x^+ -x_i^+}\partial_{-i}+
2{x^- -x_i^-\over x^+ -x_i^+}\right]
\biggl\langle h_{++}(x)\dots {h_{++}(x_i)}\dots h_{++}(x_n)\biggr\rangle\Bigr\}
\cr}
\eqn\bho
$$

$$
\biggl\langle A_+(x)A_+(x_1)\dots A_+(x_n)\biggr\rangle=4\pi Q_2^2\sum_{i=1}^n
{1\over (x^+ -x^+_i)^2}\biggl\langle A_+(x)\dots \widehat{A_+}(x_i)
\dots A_+(x_n)\biggr\rangle\ ,
\eqn\bha
$$
where we have conveniently rescaled fields as\hfill\break $h_{++}\rightarrow
(1/4\pi Q^2_1)h_{++}$,
and $A_+\rightarrow (1/\sqrt{ 2(a'-\beta^2)})2\pi Q_2
A_+$, and ``~hat~'' means omission of that term.
{}From eqs.\bho,\bha\ and \venti\ one finds the OPE for various currents
$$\eqalign{
J^a(x^+)J^b(x^+)&=-{K\over 2}{\eta^{ab}\over (x^+-y^+)^2}+{f^{ab}{}_c J^c(y^+)
\over x^+ -y^+}+ {\rm reg.}\ ,\cr
\widetilde{J}(x^+)\widetilde {J}(x^+)&=-{K_{U(1)}\over 2}{1\over (x^+-y^+)^2}+
{\rm reg.}\ ,\cr
J^a(x^+)\widetilde{J}(x^+)&={\rm reg.}\ ,\cr}
\eqn\jjj
$$
where $K$ and $K_{U(1)}$ are given by

$$
\eqalign{
K^{cl.}&=-8\pi Q_1^2\ ,\cr
K^{cl.}_{U(1)}&=-8\pi Q_2^2\ .\cr}
\eqn\ventiunouno
$$
and $f^{ab}{}_c$ and $\eta^{ab}$ are structure constants and metric of
the $SL(2,R)$ current algebra. Therefore, we see that in the case of
chiral gravity the underlying
Kac-Moody structure is $ SL(2,R)\otimes U(1)$. The solutions \venti\
allow to write the quantum version of the Sugawara
energy-momentum tensor
$$
T_{++}=-{1\over K+2}\eta^{ab}:J_a J_b:
-\partial_+ J_+ -{1\over K_{U(1)}} :\tilde J_+^2: -\partial_+
\tilde J_+\ .
\eqn\ventiuno
$$
Various OPE's of such $T_{++}$ with other fields are determined to be
$$
T_{++}(x)A_+(y)={K_{U(1)}\over(x^+-y^+)^3}+{A_+(y)\over(x^+-y^+)^2}+
{\partial_+ A_+(y)\over x^+-y^+}+{\rm reg.}\ ,
\eqn\ta
$$

$$\eqalign{
T_{++}(x)h_{++}(y)&=-{2y^- K\over(x^+-y^+)^3}+{(2h_{++}(y)-y^-\partial_-
h_{++})\over(x^+-y^+)^2}+
{\partial_+ h_{++}(y)\over x^+-y^+}+{\rm reg.}\ ,\cr}
\eqn\tipiutipiu
$$

$$
\eqalign{
T_{++}(x)T_{++}(y)&={1\over 2}{{3K\over K+2}-6K +1-6K_{U(1)}\over(x^+-y^+)^4}+
{2T_{++}(y)\over (x^+-y^+)^2}+{\partial_+T_{++}(y)\over x^+-y^+}\cr
&-{x^- -y^-\over x^+ -y^+}\left[{2\partial_-T_{++}(y)\over (x^+-y^+)^2}+
{\partial_+\partial_- T_{++}(y)\over x^+-y^+}\right]+{\rm reg.}\ ,\cr
T_{++}(x)T_{+-}(y)&={2T_{+-}(y)\over (x^+-y^+)^2}+
{\partial_+T_{+-}(y)\over x^+-y^+}+{\rm reg.}\ ,\cr
T_{+-}(x)T_{+-}(y)&={\rm reg.}\ .\cr}
\eqn\tipiumeno
$$
{}From eqs.\ta,\tipiutipiu\ we see that $A_+$ and $h_{++}$ are not primary
fields. Eqs.\tipiumeno\ show OPE of the quantum generators of the
residual symmetries in the light-cone gauge. Symmetry condition is equivalent
to the weakly vanishing of $T_{++}$ and $T_{+-}\prop J_0$ which leads to the
vanishing of the total Virasoro central charge
$$\eqalign{
&n_- -28 +c_{\rm grav.}=0\ ,\cr
&c_{\rm grav.}={3K\over K+2}-6K +1+ 48\pi Q_2^2\ .\cr}
\eqn\ventidue
$$
Comparing eq.\ventidue\ to \diciannove\
shows that these are the same equations resulting from the vanishing of the
{\it total Virasoro charge.} The difference is that in eq.\ventidue\ we
have exploited the quantum relation between Virasoro and Kac-Moody charge
$\displaystyle{c={K\cdot{\rm dim}G\over K+2}}$ [\conf], where $G$ is
appropriate symmetry group,
 and, therefore, eq.\ventidue\ is an
equation for the {\it renormalized Kac-Moody central charge.}
$$
K+2=-{1\over 12}\left[A\pm \sqrt{\left(A-12\right)\left(A+12\right)}\right]
\eqn\ventitre
$$
where $\displaystyle{A= -12-48\pi Q_1^2}$ and we have exploited the property
of the $ U(1)$ Kac-Moody central charge of being not
renormalized at the quantum level [\quant]. Eq.\ventitre\ displays {\it
dependence of the renormalized Kac-Moody central charge on the free parameter,}
while {\it Virasoro central charge of gravity is independent of such a
parameter} as visible from eq.\ventidue\ or \diciannove. The later result
has been obtained also in the conformal gauge while the former is visible
only indirectly in this gauge as explained later in the conclusions of this
letter.

Now, the reality condition
\zero\ can be written as a constraint over the allowed values $\hat a$:
$$
\left(\sqrt{n_+} - \sqrt{n_-}\right)^2<-\hat a<
\left(\sqrt{n_+} + \sqrt{n_-}\right)^2\ .
\eqn\ventitreuno
$$
Once $\hat a$ is chosen inside the range defined by \ventitreuno\
the square root in eq.\ventitre\ is real, and {\it
the renormalized $SL(2,R)$ Kac-Moody central
charge is real too, without any restriction over $n_\pm $.} This is our main
conclusion.
The above result has fulfilled our hope that the presence of Lorentz
anomaly (~and therefore of a free parameter~) improves the value of the
renormalized Kac-Moody central charge with respect to the non-chiral
gravity, where exist regions of complex values of the renormalized
Kac-Moody charge.
It is therefore reasonable to expect that the physical
parameters such as conformal dimensions and  string susceptibility will
depend on the free parameter as well. The latter is expressed in terms of
the renormalized Kac-Moody central charge by:
$\Gamma=K+3$. If we use eq.\ventitre\ to compute $\Gamma$, we reproduce
a result previously found in the conformal gauge [\meyer].

In this letter we have considered the Kac-Moody structure of the chiral
induced gravity in the light-cone gauge. We have shown that the presence
of the Lorentz degree of freedom produces results different from non-chiral
situation. First of all, Kac-Moody algebra is enlarged by the Lorentz $U(1)$
factor; secondly, renormalized $SL(2,R)$ Kac-Moody central charge has no
forbidden regions thanks to the presence of an arbitrary parameter. Our results
agree with those obtained in conformal gauge by the DDK method [\dk]. Although
comparison between the two gauges is  possible, as in [\kl], $SL(2,R)$
symmetry is not directly visible in the conformal gauge but only through
the comparison of the expressions for
invariant quantities such as string susceptibility and scaling
dimensions. Details of this comparison will be given elsewhere, but we
anticipated the above relation for the string susceptibility proving
agreement to the conformal gauge.
The role of the regularization parameter has also been considered in
[\oz], where it has been claimed that the light-cone analysis
selects the value $\hat a=-(n_+ -n_-)/192\pi$
for the regularization parameter. This choice does not seem to be
justified neither in our analysis, nor
in conformal gauge. The result proposed in ref.[\oz] is equivalent,
in our language, to setting
$Q_2=0$. But, we find no physical reason for such a choice.

Finally, as far as the presence of a ghost-like field is concerned, we recall
that a mechanism to decouple the Lorentz
ghost from the physical sector has been proposed for the model
in the conformal gauge [\lee]. It is based on the usual treatment of
ghosts in the BRST formalism.
The same decoupling mechanism works in the light-cone gauge as well,
therefore resolving  the  problem of unitarity in this model.

\refout
\bye